\documentclass[aps,prx,superscriptaddress,twocolumn,showpacs,showkeys,floatfix]{revtex4-1}
\usepackage{amsmath}
\usepackage{bm}
\usepackage{amssymb}
\usepackage{graphicx}
\usepackage{epstopdf}
\usepackage{epsfig}
\usepackage{color}
\usepackage{hyperref}
\usepackage{float}
\usepackage{braket}
\usepackage[mathscr]{euscript}
\begin{document}

\title{Controlling electric, magnetic, and chiral dipolar emission with PT-symmetric potentials}
\author{Hadiseh Alaeian}
\affiliation{Department of Electrical Engineering, Stanford University, Stanford, California 94305, USA}
\affiliation{Department of Materials Science and Engineering, Stanford University, Stanford, California 94305, USA}
\author{Jennifer A. Dionne}
\affiliation{Department of Materials Science and Engineering, Stanford University, Stanford, California 94305, USA}
\date{\today}

\begin{abstract}
We investigate the effect of parity-time (PT)-symmetric optical potentials on the radiation of achiral and chiral emitters. Mode coalescence and the appearance of exceptional points lead to orders-of-magnitude enhancements in the emitted dipole power. Further, the emitter can be tuned to behave as a strong optical source or absorber based on the non-Hermiticity parameter. Chiral enantiomers radiating near PT metamaterials exhibit a 4.5-fold difference in their decay rate. The results of this work could enable new atom-cavity interactions for quantum optics, as well as all-optical enantio-specific separation.
\end{abstract}

\maketitle 

\section{introduction}
The rate of spontaneous emission from a quantum system is not an intrinsic property.  Instead, an emitter's radiative rate and decay time can be significantly influenced by its surroundings.  Since the pioneering work of Purcell on cavity-emitter interactions~\cite{Purcell64}, considerable research has explored new materials and geometries to enhance radiative rates, including photonic crystals ~\cite{Noda07,Iwase10,Canet-Ferrer12,Canet-Ferrer13}, plasmonic structures ~\cite{Koenderink10,Ma11,Shubina10,Vesseur10} and metamaterials ~\cite{Jacob12,Poddubny11,NPoddubny13,Poddubny13}. Each of these systems tailors light-matter interactions by modifying the local density of optical states (LDOS), which in turn dictates the number of radiative and non-radiative pathways available to an emitter for decay. 

Recently, parity-time (PT) symmetric potentials have offered a new platform to tailor light-matter interactions. These potentials rely on the balanced inclusion of loss and gain media, and render the optical Hamiltonian non-Hermitian. Below a so-called `exceptional point', PT-symmetric systems will be characterized by a real eigenspectrum despite their non-Hermiticity ~\cite{Bender98,Bender04,Bender005,Bender05}. Thereafter, eigenvalues will move into the complex plane and become complex conjugates of each other. Accordingly, optical modes can propagate preferentially in one spatial location or another, exhibiting either optical gain or strong attenuation ~\cite{El-Ganainy07, Guo09, Makris08,Makris10,Ruter10, Alaeian14}. The unique and unidirectional optical properties attainable with PT potentials has enabled applications ranging from optical diodes and insulators to laser-absorbers ~\cite{Chong11,Liu14,Longhi14}. 

While the interaction of plane waves with PT-symmetric media has been well-studied, the spontaneous emission of quantum emitters near PT potentials remains unexplored. In this work, we investigate the radiation of electric and magnetic dipoles near PT-symmetric metamaterials. We begin by exploring achiral emitters, showing how both the magnitude and sign of the radiated power can be tuned. Depending on the strength of the PT potential (i.e., the `non-Hermiticity parameter'), the emitter can act as a strong optical source or an efficient absorber, with positive or negative Purcell factors.  Further, the radiative rate can be increased by several orders of magnitude at the exceptional point, where the eigenstates coalesce and increase the LDOS. Subsequently, we explore the radiation of chiral emitters near PT metamaterials. Through appropriate design of PT-symmetric potentials, we show how enantiomers can be distinguished by their decay rate, with maximum differences observed at the exceptional point. Coupled with a photoionization scheme to selectively target excited-state molecules, as proposed in ~\cite{Klimov12}, these results could facilitate efficient optical enantiomer separation.

\section{Theoretical Formulation}

\begin{figure}
\includegraphics[scale=0.9]{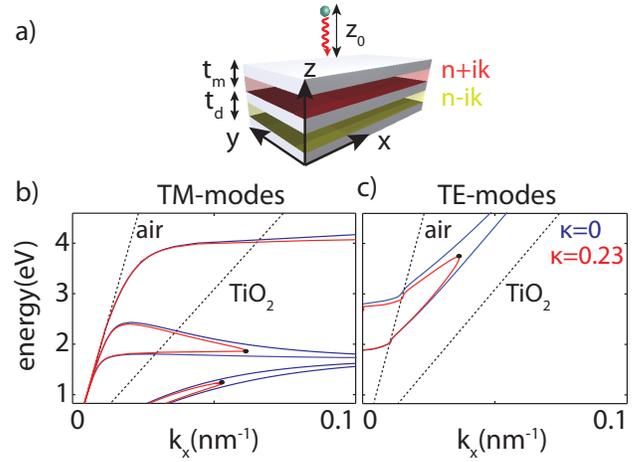}
\caption{\label{fig1}(a) Schematic of a dipole radiating in the vicinity of the 5-layer PT-symmetric metamaterial. Dispersion curves of the (b) TM modes and (c) TE modes of the metamaterial for two values of the non-Hermiticity parameter, $\kappa=0$ (blue) and $\kappa=0.23$ (red). The black circles denote the exceptional points. The dashed lines correspond to the light lines in the air and dielectric.}
\end{figure}

We consider the planar plasmonic metamaterial shown in Fig.~\ref{fig1} (a), composed of a five-layer stack of alternating layers of metal and dielectric. The layers are assumed to be infinite in the $xy$-plane but finite in $z$. The metal and dielectric thicknesses, $t_m$ and $t_d$, are deeply subwavelength and taken to be 30 nm.  The metal is modeled as a lossless Drude material with a permittivity $\epsilon=1-(\frac{\omega_p}{\omega})^2$. The plasma frequency, $\omega_p$, is taken to be $8.85\times 10^{15} s^{-1}$, similar to bulk plasma frequency of Ag. The dielectric layers have a refractive index $n\pm i\kappa$, with one layer corresponding to loss media (+$\kappa$) and the other corresponding to gain media (-$\kappa$). For concreteness, we consider n=3.2, corresponding to the refractive index of TiO$_2$ in the frequency range of interest. The imaginary part of the refractive index $\kappa$ is variable, but it is always identical in each dielectric layer to satisfy the PT-symmetric condition. The dipole emitter is assumed to be a distance $z_0$ away from the first vacuum/metal interface of the structure. 

Fig.~\ref{fig1} shows the dispersion curves for the metamaterial, indicating that both transverse magnetic (TM) and transverse electric (TE) modes are supported. Each panel includes calculations for two values of the non-Hermiticity parameter, $\kappa$=0 and $\kappa$=0.23. At $\kappa=0$, the in-plane wave vector $k_x$ diverges for TM modes (Fig.~\ref{fig1} (b)) at the Ag-TiO$_2$ and Ag-vacuum surface plasmon resonance frequencies (E = 1.7 eV and 4 eV, respectively). Wavevectors remain finite and smaller than the TiO$_2$ light line for TE modes. As the non-Hermiticity parameter is increased, modes converge toward the same energy and wave vector, and coalesce at the exceptional points (EP), denoted by black circles. This point is of particular importance as it shows a phase transition in the modal behavior of the waveguide. Before this EP, the modes have real propagation constants and field distributions have a definite symmetry. After the EP however, the propagation constants move into the complex plane and the fields lose their symmetry. This region beyond the exceptional phase is called the `broken phase.' As described in reference ~\cite{Alaeian14B}, in the broken phase, one mode is localized almost exclusively in the gain media, while the other is confined to the loss media. 

To determine how these metamaterial modes impact dipolar emission, we calculate the Purcell factor, defined as the power radiated by a dipole, $P$, normalized to its radiated power in free space $P_0$ ~\cite{Novotny06}: 

\begin{equation}\label{PF-eq}
\begin{split}
\frac{P}{P_0}=1+\frac{3}{4}\frac{|\vec{p}_\rho|^2}{|\vec{p}|^2}\int\limits_{0}^{\infty}Re[\frac{k_\rho}{k_z}(r_{TE}-r_{TM} k_z^2)e^{i2k_zz_0}]dk_\rho\\
+\frac{3}{2}\frac{|p_z|^2}{|\vec{p}|^2}\int\limits_{0}^{\infty}Re[\frac{k_\rho^3}{k_z}r_{TM} e^{i2k_zz_0}]dk_\rho
\end{split}
\end{equation}

Here, $\vec{p}$, $\vec{p}_\rho$ and $p_z$ denote the electric dipole moment and its transverse and normal components, respectively. Likewise, $k_\rho$ is the transverse momentum in the $xy$-plane ($k_\rho=\sqrt{k_x^2+k_y^2}$), and $r_{TE}$ and $r_{TM}$ are the reflection coefficients from the metamaterial for TE- and TM-polarizations. 

In general, this equation implies three important features of dipolar emission near a PT plasmonic metamaterial. Firstly, the Purcell factor strongly depends on the modal wave vector and hence momentum. Therefore, at the surface plasmon resonance frequencies where mode momenta diverge, the LDOS increases and a significant modification of the Purcell factor is expected. Secondly, the Purcell factor strongly depends on the reflection coefficient. As discussed in the next section, the reflection coefficient can be modified with increasing the non-Hermiticity parameter. An abrupt change in the divergence of the reflection coefficient at the exceptional point noticeably enhances the Purcell factor (Appendix B details the behavior of the S-matrix poles). Lastly, eq.~\ref{PF-eq} suggests that the reflection coefficient can control the sign of Purcell factor as well. As shown in Appendix A, the reflection coefficients of evanescent components ($k_\rho\ge k_0$) interacting with the gain or loss side of PT media are always complex conjugate of each other: $r_G=r_L^*$. For these evanescent components, $k_z$ is purely imaginary, thus the exponential term $e^{i2k_zz_0}$ is real and the power spectrum is directly proportional to the imaginary part of the reflection coefficients. Accordingly, the non-radiative power changes sign when the reflection coefficient is replaced with its complex conjugate - or physically, when a dipole is repositioned from the loss to the gain side. Ultimately, whenever the non-radiative contribution is dominant (i.e. when the dipole is close to the structure), this feature can change the sign of the total power $P$. This intriguing result complements the reports of asymmetric reflections of propagating plane waves from PT structures when illuminated from the loss of gain side ~\cite{Alaeian14}. In the following sections, we present the numerical results particular to the structure depicted in Fig.~\ref{fig1}.

\section{achiral emitter}
\begin{figure}
\includegraphics[scale=0.89]{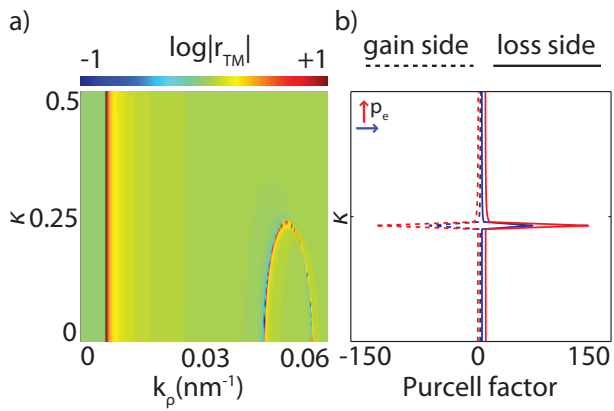}
\caption{\label{fig2} (a) Reflection coefficient of TM-polarized planewave as a function of in-plane momentum and non-Hermiticity parameter $\kappa$. The energy of the planewaves is 1.2 eV. (b) Purcell factor of a vertical (red lines) and horizontal (blue lines) electric dipole as a function of $\kappa$ at E=1.2 eV and $z_0$=20 nm. The solid lines show the total emitted power when the dipole is close to the loss layer while the dashed lines correspond to the gain side.}
\end{figure}

Since the power emitted by a dipole is directly related to the reflected fields, we started by investigating the reflection coefficients. Figure~\ref{fig2}(a) plots the variation of the reflection coefficient with non-Hermiticity $\kappa$ parameter and in plane momentum $k_\rho$. We consider TM-polarized illumination, and set the energy to E=1.2eV. 
At this energy, all modes supported by the metamaterial lie below the light line and have real momenta exceeding that of free space (refer to Fig.~\ref{fig1}(b)). As seen, the reflection coefficient diverges for wavevectors corresponding to the guided modes. For $\kappa=0$, this divergence occurs for three wavevectors  ($k_\rho=$0.006, 0.046, and 0.058 nm$^{-1}$).  As the non-Hermiticity parameter is increased, the lowest wave vector mode exhibits minimal variation. However, the higher-momenta modes have reflection coefficients that begin to coalesce and form a loop in $k_\rho \kappa$-plane, terminating at the exceptional point ($\kappa=0.23$). For larger values of $\kappa$, the reflection coefficient at these larger wavevectors decreases, due to the lack of momentum matching between guided modes and incident planewaves.  A similar study on the reflection coefficient of TE-modes leads to a featureless map, due to the lack of TE-modes at this low energy. (see Appendix C).

Figure~\ref{fig2}(b) shows the total power radiated by an electric dipole located 20 nm away from the metamaterial. We consider both horizontal and vertical dipoles at an energy of 1.2 eV. As seen, the Purcell factor increases by two orders of magnitude at the exceptional point. This behavior is nearly independent of dipole orientation, with slightly more power observed for the vertical dipole as it completely couples to TM modes. Also, Fig.~\ref{fig2}(b) indicates that the sign of the total power changes based on whether the dipole is located on the gain side (dashed lines) or loss side (solid lines) of the metamaterial. As described before, the non-radiative part of the power spectrum experiences complex conjugated reflection coefficients from the gain and loss side. This implies that the non-radiative part of the power changes sign as the dipole is relocated from the gain side to the loss side. Here, the dipole's close proximity to the interface means that the non-radiative contribution dominates the radiative contribution by about two-orders of magnitude. Therefore if the sign of the non-radiative part is changed, the sign of the total power can also be changed. While the large positive Purcell factor from the loss side means that the dipole behaves as an efficient emitter, the negative sign on the gain side implies that the dipole efficiently absorbs power.

\begin{figure}
\includegraphics[scale=0.89]{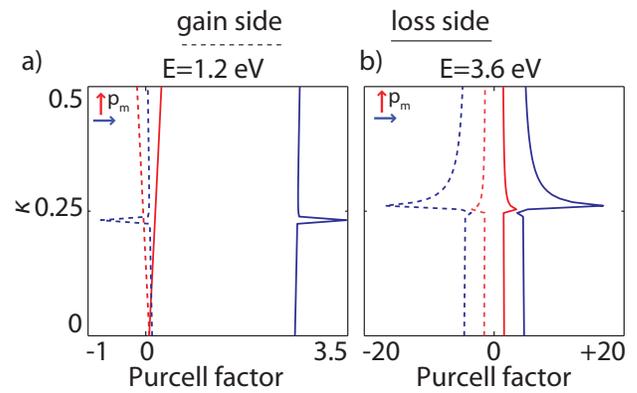}
\caption{\label{fig3} Normalized power emitted by a vertical (red) and a horizontal (blue) magnetic dipole as a function of $\kappa$ at (a) E=1.2 eV and (b) E=3.6 eV. The dipole is assumed to be 20 nm away from the first interface. The solid lines show the emitted power for a dipole close to the loss while the dashed lines correspond to the dipole close to the gain side.}
\end{figure}

We also calculate the emitted power from magnetic dipoles. Figure~\ref{fig3}(a) plots the Purcell factor for both horizontal and vertical magnetic dipoles as a function of $\kappa$ at E=1.2eV. Here, unlike electric dipolar emitters, significant differences are observed between dipole orientations. While the horizontal magnetic dipole shows a maximum at the exceptional point ($\kappa\approx 0.23$), the vertical magnetic dipole has no resonance feature. A horizontal magnetic dipole excites both TE and TM-polarizations while a vertical dipole exclusively couples to TE-modes. As shown in Fig.~\ref{fig1}(c) the structure supports no TE mode at this low energy, hence no exceptional point will be observed at E=1.2 eV for TE-modes. Accordingly, Purcell factors remain small for vertically-oriented dipoles. Further, note that the total power for horizontal magnetic dipoles is not symmetric. This asymmetry is a general feature for all dipoles near PT media, but is magnified for this particular case since the ratio between non-radiative and radiative contributions is small.  While the non-radiative part still contributes dominantly to the total power, it only is about three times larger than the radiative part.

At higher energies, this structure can support both TM and TE modes. For example, at E=3.6 eV, the TE-reflection coefficient in the $k_\rho \kappa$-plane shows a similar loop at $\kappa=0.26$ (see Appendix C). Therefore, unlike E=1.2 eV, at 3.6 eV both TE and TM modes exhibit exceptional points in their spectra. Figure~\ref{fig3}(b) shows the total power radiated by both vertical and horizontal magnetic dipoles at this energy. Local maxima in the Purcell factor are observed for all four configurations. In particular, notice that the vertical magnetic dipole, which exclusively couples to TE-modes, has a resonant peak at $\kappa$ =0.26, corresponding to the exceptional point of these modes at this energy.

Figures 2 and 3 imply that mode coalescence at the exceptional points significantly modifies the power dissipation spectrum (the integrand of eq.~\ref{PF-eq}) and the total power. The poles of the reflection coefficients (or S-matrix) provide a deeper understanding of this phenomena. Before the exceptional point, the two simple poles, corresponding to the two metamaterial modes, contribute oppositely to the integral and hence the total power. At the exceptional point, the modes coalesce to form a double pole, so this opposite behavior vanishes. Therefore, a marked increase in the integral and power is obtained. After the exceptional point, only one simple pole contributes. However the contribution of this pole monotonically decreases as the pole moves away from the real axis (larger $\kappa$), hence the total power again decreases. Further details can be found in Appendix B. 

\begin{figure}
\includegraphics[scale=0.89]{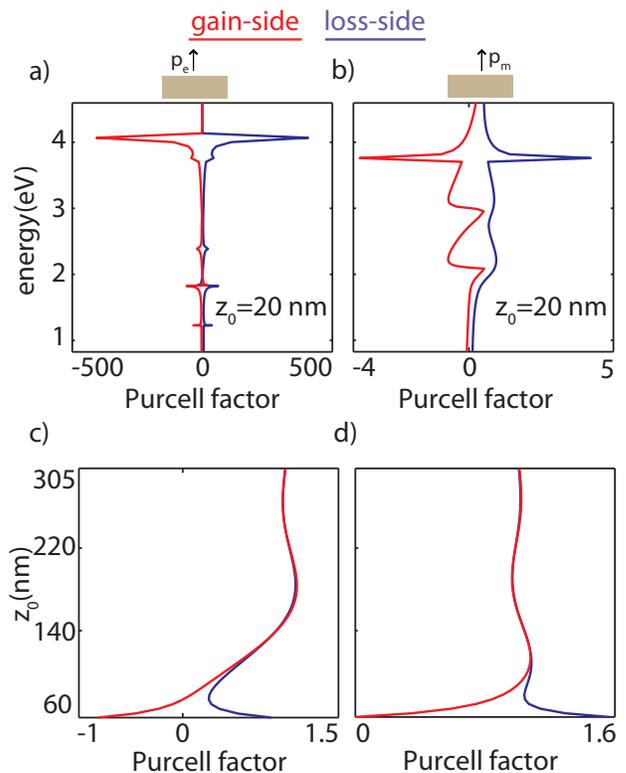}
\caption{\label{fig4}Purcell factor of (a) a vertical electric dipole and (b) a vertical magnetic dipole near the PT-symmetric metamaterial as a function of energy. In both cases $\kappa=0.23$ and $z_0$=20 nm. The variation of the Purcell factor as a function of $z_0$ at E=4 eV and $\kappa$=0.23 is shown for (c) a vertical electric dipole and (d) a vertical magnetic dipole. Note that in these two figures the lower limit is increased to $z_0$= 60 nm for better illustration.}
\end{figure}

The spectral variation of the radiated power is shown in Fig.~\ref{fig4} plots. Both vertical electric and magnetic dipoles are included. As seen in (a) and (b), which consider a dipole positioned 20 nm above the metamaterial, peaks in the Purcell factor appear at both exceptional point frequencies and surface plasmon resonance frequencies. For example, a vertical electric dipole couples exclusively to TM modes and exhibits local maxima in the Purcell factor at energies of 1.2 eV and 1.9 eV (the exceptional points for the four lowest order branches) and at 2.3 eV and 4 eV. In contrast, magnetic dipole radiation cannot couple to TM modes at E=1.2 eV. However, its power spectrum has a resonance feature at E=3.8 eV, where an exceptional point arises for $\kappa$=0.23. Variation of the Purcell factor at lower energies appears due to the cut-off of various TE modes around 1.9 and 2.7 eV.

The relative contribution of radiative and non-radiative components to the Purcell factor varies strongly as a function of $z_0$ (the dipole-metamaterial separation). While the non-radiative component exponentially decreases with separation, the radiative part oscillates. Since the non-radiative contribution can change the sign of the total power, the sign can in turn modified with dipole-metamaterial separation.  Figure\ref{fig4}(c) plots the spatial variation of dipole power for a z-oriented electric dipole. When the dipole is close to the structure, the power is positive on the loss side and negative on the gain side. For larger separations (z$\geq$78 nm), the power radiated from the dipole is always positive, independent of its proximity to the gain or loss side. As the separation approaches infinity, the Purcell factor approaches unity, as expected. Similar trends hold for magnetic dipoles, though the magnetic dipole needs to be placed within 60 nm of the metamaterial to obtain a similar change in sign. Accordingly, a dipole located on the gain side can be tuned to behave as a bright emitter (positive power) to an efficient absorber (negative power) by changing its separation.

\section{chiral emitters}
The emergence of chirality is largely attributed to the interaction of simultaneous electric and magnetic dipoles ~\cite{Barron82, Klimov05}. Consequently, as with a chiral emitters, the decay rate and radiated power of chiral molecules can be modified with the environment. Recently, the interaction of chiral and achiral molecules with chiral objects has been the subject of extensive study ~\cite{Lakhtakia90, Klimov14, Guzatov12, Klimov12}. It has been shown that enantiomers exhibit enantio-specific coupling to the modes of a chiral scatterer, and that chiral structures can substantially modify the decay rate and radiation pattern of chiral molecules ~\cite{Klimov12}. Here, we consider the radiation of a chiral molecule in the vicinity of our PT-symmetric structure, which contains no chiral constituents. 

Equation ~\ref{PF-eq} can be extended to include the simultaneous radiation of the electric and magnetic dipoles. Doing so, the normalized power radiated by a chiral source is given by:

\begin{equation}
\frac{P}{P_0}=1+\frac{\omega}{2P_0}Im[\vec{p}_e^*\cdot\vec{E}^s(\vec{r}_0)+\vec{p}_m^*\cdot\vec{B}^s(\vec{r}_0)]
\end{equation}

In this equation, $\vec{p}_e$ and $\vec{p}_m$ are the electric and magnetic dipole moments of the molecule, while $\vec{E}^s$ and $\vec{B}^s$ are the scattered electric and magnetic fields at the position of the molecule, $\vec{r}_0$. $P_0$ is the power radiated by a chiral source in free space. It can be shown that $P_0$ is given by the summation of the power emitted by each dipole in free space individually. Since the magnetic moment operator is purely imaginary for a two-level system, a $\pi/2$ phase difference exists between the electric and magnetic dipoles. More complex chiral molecules are characterized by a variable phase relationship (and the possible need for quadrupolar terms). For simplicity, we only consider dipolar terms here. We use the common naming convention based on the sign of $\vec{p}_e\cdot\vec{p}_m$, where a right-handed enantiomer refers to a positive product, while a left-handed enantiomer refers to a negative dot product.  

\begin{figure}
\includegraphics[scale=0.9]{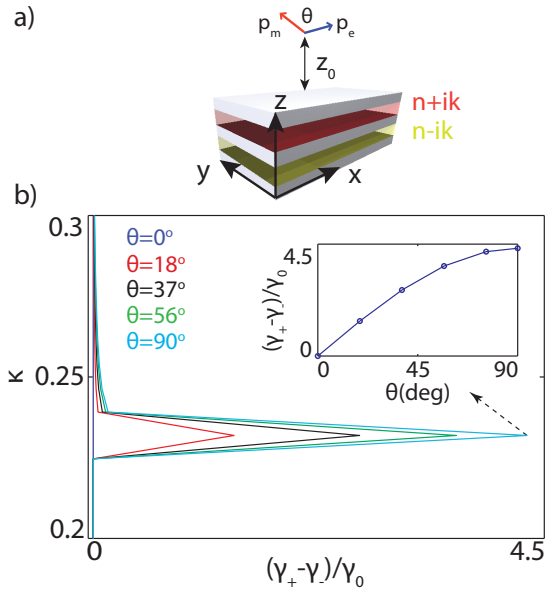}
\caption{\label{fig5}(a) Schematic of a chiral molecule composed of electric and magnetic dipoles separated by angle $\theta$. (b) The difference between the decay rate of right (+) and left (-) enantiomers as a function of $\kappa$ for different angles. The energy of the emitter is 1.2 eV. The inset shows the variation of the peak value as a function of the angle $\theta$ between the two dipoles.}
\end{figure}

A schematic of a chiral molecule close to our metamaterial is shown in Fig.~\ref{fig5}(a). The electric and magnetic dipoles are located 20 nm away from the interface and in the $x$-$y$ plane with an angle $\theta$ between them. Also, based on references~\cite{Klimov12,Klimov12,Klimov14}, we assume that the ratio of the electric and magnetic dipoles is $\xi=0.1$. From Fig.~\ref{fig3} and ~\ref{fig4}, we know the chiral molecule will exhibit an increased Purcell factor and variable sign near the PT metamaterial. Is it possible to distinguish enantiomers based on their radiation near this metamaterial?

For chiral selectivity, there must be an effect from the electric dipole at the position of the magnetic dipole and vice-versa. Otherwise, due to the sign relationship of $\vec{p}_e\cdot\vec{p}_m$ for the two enantiomers, there would be no change in the total power radiated by each enantiomer. Figure\ref{fig5}(b) plots the difference between the decay rates of the right ($'+'$) and left ($'-'$) enantiomers as a function of $\kappa$. The energy is fixed at 1.2 eV. The parameters for the left enantiomer have been calculated by substituting $\vec{p}_m$ with $-\vec{p}_m$, while $\vec{p}_e$ is always fixed along the $x$-direction. While the difference between decay rates is minimal below the exceptional point, at this exceptional point the decay rates are markedly different. This difference monotonically increases by increasing the angle between the dipoles. Note that an $x$-directed electric dipole at $r_0$ produces only non-zero $H_y$ at this point. Therefore, as the angle between the dipoles approaches $90^o$, the scattered magnetic field by an electric dipole increases. At $\theta=90^o$, the difference between enantiomer decay rate is maximized to 4.5. In other words, if a racemic mixture of chiral enantiomers are excited, the right enantiomer decays 4.5 times faster than the left enantiomer to its ground state. If combined with a photo-ionization technique to remove molecules in the excited state, this interaction could be used to form an enantiopure product. 

\section{conclusion}
We have studied the effect of a PT-symmetric optical potential on the radiation of achiral and chiral molecules. PT-symmetric potentials not only tune the value of the Purcell factor but also the sign. For simple electric or magnetic dipoles, mode coalescence at the exceptional point increases the radiative power by orders of magnitude. Further, the broken phase allows for a change in the sign of the radiated power. A dipole can serve as a bright emitter or an efficient absorber based on its position with respect to the metamaterial (loss or gain side), and also its height above the metamaterial. Further, the exceptional point leads to a 4.5x difference in left/right enantiomer decay rates. Looking forward, these results could be utilized in the design of new PT-symmetric cavities to control emitter properties. For example, the large Purcell factors at exceptional points could change a normally 'dark' molecule a bright emitter; or, alternatively, a bright emitter could be switched to an efficient absorber by re-locating the dipole. Such effects could be utilized in designing an efficient all-optical, single-photon modulator or a sensitive molecular ruler. The results might also pave the way for all-optical enantio-selective separation.

\section*{Acknowledgements}
The authors greatly appreciate useful feedback from all Dionne group members, in particular Dr. Gururaj V. Naik, Dr. Yang Zhao and Dr. Aiztol Garcia-Etxarri. Funding from a Presidential Early Career Award administered through the Air Force Office of Scientific Research is gratefully acknowledged, as are funds from Northrop Grumman.
Funding from the Hellman Faculty Scholars program and a National Science Foundation CAREER Award (DMR-1151231) are also gratefully acknowledged.

\appendix
\section{Scattering properties of a PT-symmetric potential}
The behavior of a multilayered structure can be described using either a transfer matrix \emph{T} or scattering matrix \emph{S}. The transfer matrix favours itself to cascaded systems via the multiplication of each layer $T$ matrix as: 
\begin{equation}
T_{eq}=\prod\limits_{n=1}^{n=N}T_n
\end{equation} 
where the $m^{th}$ layer transfer matrix is given by:
\begin{equation}
\begin{split}
T_m=[air]^{-1} I_m D_m I_m^{-1}[air]\\
I_m=\begin{pmatrix}
1 & 1\\
\frac{k_m}{\alpha_m} & -\frac{k_m}{\alpha_m}
\end{pmatrix}\\
D_m=\begin{pmatrix}
e^{ik_md_m} & 0\\
0 & e^{-ik_md_m}
\end{pmatrix}\\
[air]=\begin{pmatrix}
1 & 1\\
k_{air} & -k_{air}
\end{pmatrix}
\end{split}
\end{equation}
where $k_m=\sqrt{k_0^2\epsilon_m-k_x^2}$ and $\alpha_m=1$ and $\epsilon_m$ for TE and TM polarizations, respectively.
From these equations the total transfer matrix can be written as:
\begin{equation}
T_{eq}=[air]^{-1}(I_ND_NI_N^{-1})\cdots(I_2D_2I_2^{-1})(I_1D_1I_1^{-1})[air]
\end{equation}
Therefore the following equation gives the inverse of the transfer matrix as:
\begin{equation}
T_{eq}^{-1}=[air]^{-1}(I_1D_1^{-1}I_1^{-1})(I_2D_2^{-1}I_2^{-1})\cdots(I_ND_N^{-1}I_N^{-1})[air]
\end{equation}
if $\epsilon_m\rightarrow\epsilon_m^*$ then $I_m\rightarrow I_m^*$ and $D_m\rightarrow D_m^{*-1}$.

In a PT-symmetric potential, the permittivity distribution satisfies $\epsilon(z)=\epsilon^*(-z)$. Hence, the spatially symmetric layers either have the same real refractive indices or the permittivities are complex conjugate of each other. Assume that layer \emph{m} and \emph{N-m}+1 have complex conjugated permittivities. Therefore we have:
\begin{equation}
\begin{split}
I_m=I_{N-m+1}^*\\
D_m^{-1}=D_{N-m+1}^*\\
I_m^{-1}=I_{N-m+1}^{*-1}
\end{split}
\end{equation} 
hence:
\begin{equation}
T_m^{\prime -1}= T_{N-m+1}^{\prime *}
\end{equation}
where
\begin{equation}
T_m^\prime=I_mD_mI_m^{-1}
\end{equation}

If the $m^{th}$ layer is lossless (i.e., has a real refractive index), then $k_m$ can be either a pure real or pure imaginary number.

\emph{case 1}: $k_m$ is real:
\begin{equation}
\begin{split}
I_m=I_m^*\\
D_m^{-1}=D_m^*\\
I_m^{-1}=I_m^{-1*}
\end{split}
\end{equation}
hence $T_m^{\prime -1}=T_m^{\prime *}$.

\emph{case 2}: $k_m$ is imaginary:
\begin{equation}
\begin{split}
T_m^\prime=\frac{\alpha_m}{2k_m}\begin{pmatrix}
1 & 1\\
\frac{k_m}{\alpha_m} & -\frac{k_m}{\alpha_m}
\end{pmatrix}
\begin{pmatrix}
e^{+ik_md_m} & 0\\
0 & e^{-ik_md_m}
\end{pmatrix}
\begin{pmatrix}
\frac{k_m}{\alpha_m} & +1\\
\frac{k_m}{\alpha_m} & -1
\end{pmatrix}\\
=\frac{\alpha_m}{2k_m}\begin{pmatrix}
2\frac{k_m}{\alpha_m} cos(k_md_m) & i2 sin(k_md_m)\\
i2(\frac{k_m}{\alpha_m})^2 sin(k_md_m) & 2\frac{k_m}{\alpha_m} cos(k_md_m)
\end{pmatrix}
\end{split}
\end{equation}

Note that in this case $k_m^*=-k_m$ hence again $T_m^{\prime -1}=T_m^{\prime *}$.
Now, rewrite the transfer matrix in the following form:
\begin{equation}
T_{eq}=[air]^{-1}A[air]
\end{equation}
where $A^{-1}=A^*$ and $|A|=1$. Therefore $A$ has the following general form:
\begin{equation}
A=\begin{pmatrix}
a & ib\\
ic & a^*
\end{pmatrix}
\end{equation}
Also $[air]$ is given as:
\begin{equation}
[air]=\begin{pmatrix}
1 & 1\\
\sqrt{k_0^2-k_x^2} & -\sqrt{k_0^2-k_x^2}
\end{pmatrix}
\end{equation}
If the waves are propagating in air, where $k_x\leq k_0$ then $[air]=[air]^*$. In this case the total transfer matrix satisfies the property of $T_{eq}^*=T_{eq}^{-1}$.

However, when the waves are evanescent, i.e. $k_0\leq k_x$, this equality no longer holds. The general form for the transfer matrix is given as:
\begin{equation}
T=\frac{1}{2\gamma}\begin{pmatrix}
2\gamma Re(a)+(c-b\gamma^2) & +i2\gamma Im(a)+(c+b\gamma^2)\\
+i2\gamma Im(a)-(c+b\gamma^2) & 2\gamma Re(a)-(c-b\gamma^2)
\end{pmatrix}
\end{equation}
where $\gamma=\sqrt{k_x^2-k_0^2}$.

Although this matrix does not satisfy the previous condition of $T^*=T^{-1}$, it leads to the new equality of $r_L=r_R^*$. In other words, evanescent planewaves are reflected with complex conjugated coefficients from the gain and loss sides of a PT-symmetric potential.

\section{Effect of the poles on the emitted power}
\begin{figure}
\includegraphics[scale=0.9]{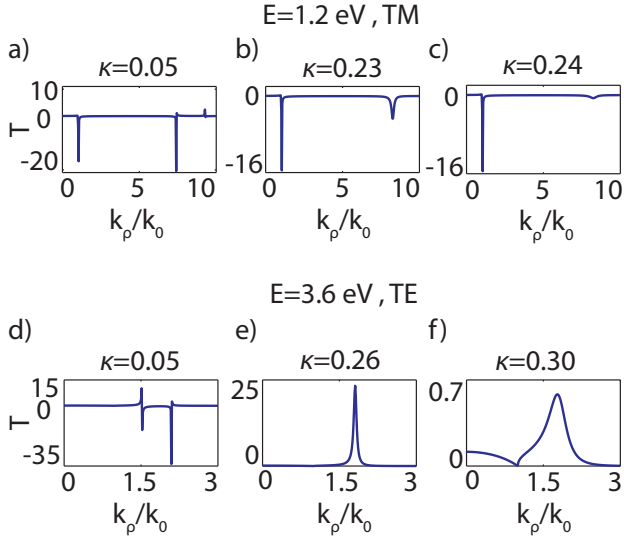}
\caption{\label{fig6} Variation of the transmission coefficient as a function of in-plane momentum and various values of $k$ for (a)-(c) TM modes at E = 1.2 eV and (d)-(f) TE modes at 3.6 eV. The non-Hermiticy values in each case are chosen to be before, at, and after the exceptional point.}
\end{figure}

In the main text, we ascribed changes in dipolar radiation near PT potentials to changes of the reflection coefficient with increasing non-Hermiticity. Here, we more quantitatively describe changes the reflection coefficients, based on the poles of the S-matrix. The poles of the S-matrix correspond to the modes of the system, and are identical to the poles of the reflection and transmission coefficients.

\begin{figure}
\includegraphics[scale=0.9]{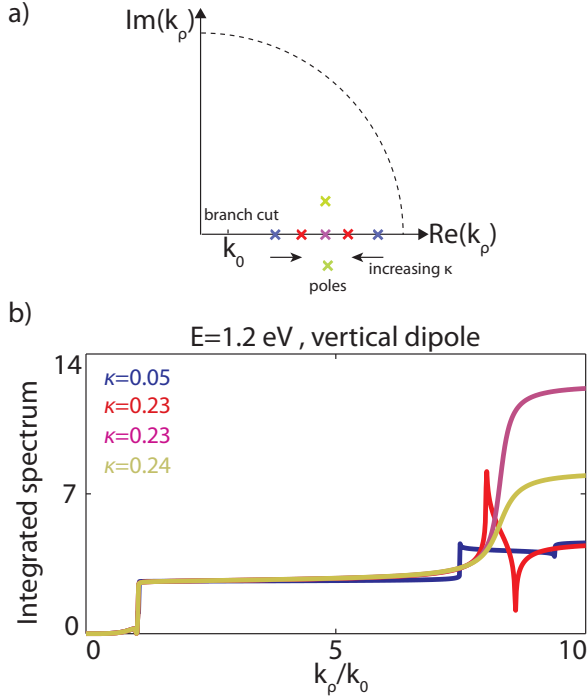}
\caption{\label{fig7} (a) Location and the poles in $k_\rho$-plane and effect of the non-Hermiticy parameter in moving the poles. (b) Partially integrated power spectrum of a vertical electric dipole at E=1.2 eV as a function of endpoint for different values of $\kappa$.}
\end{figure}

Figure~\ref{fig6} shows the transmission coefficients of the 5-layer metamaterial as a function of $k_\rho$ for three distinct values of non-hermiticity parameter $\kappa$. Aside from the common feature of branch cut at $k_\rho=k_0$ shown in Fig.~\ref{fig6}(a)-(c), $t_{TM}$ shows very different behavior for various values of $\kappa$. Referring to the dispersion diagrams of this waveguide, at E=1.2 eV the waveguide supports two deeply sub-wavelength TM-modes. The wavevector of these modes corresponds to the divergence of $t_{TM}$ in Fig.~\ref{fig6}(a) (note the sharp resonance features in this panel at larger values). This divergence is a generic behavior of a simple pole and corresponds to a non-degenerate mode in the structure. The transmission in the vicinity of the $n^{th}$ simple pole can be approximated as $\frac{A_n}{k_\rho-k_n}$. Note that this function changes its sign around the pole $k_n$. However, the different zero-crossings of the transmission close to these poles implies the change in the pole signs. More specifically, while the first pole has a negative to positive zero crossing the other pole has a positive to negative crossing. Therefore, the power dissipation spectrum (the integrand of Eq.~\ref{PF-eq}) has simple poles with opposite residues at these points. Thus although the structure supports two modes at E = 1.2 eV, these modes contribute oppositely to the total power of Eq.~\ref{PF-eq}. In other words the residue of the power dissipation spectrum at each of these simple poles $A_n$ has different signs.

As $\kappa$ increases, the poles corresponding to the modes of the structure, approach each other and finally coalesce at $\kappa=0.23$, as seen in Fig.~\ref{fig6}(b). Notice that there is no sign change around this pole and the value is exclusively negative, a signature of a second order mode and state coalesce. Accordingly, the power dissipation spectrum around this pole can be approximated as $\frac{A_n}{(k_\rho-k_n)^2}$. Increasing $\kappa$ beyond this point leads to a significant decrease in the transmission due to the new location of the poles in the complex plane and away from the real axis. A similar behavior for the TE polarized modes at E=3.6 eV has been shown in Fig.~\ref{fig6} where a comparison before, at, and after the exceptional point is given in panels (d),(e) and (f), respectively. Again, note how the two sharp resonance features accompanied by a sign change for simple poles in Fig.~\ref{fig6}(d) is substituted with a single-valued single peak at the exceptional point in Fig.~\ref{fig6}(e). Also note that the peak drastically decreases in Fig.~\ref{fig6}(f) where the poles have imaginary values after the exceptional point.

To numerically test the results, Fig.~\ref{fig7} shows the partial integral of the power dissipation spectrum. The upper limit of the integral in Eq.~\ref{PF-eq} is replaced with a variable $k_\rho$. The power dissipation integral is for a vertical dipole radiating at E=1.2 eV. Figure~\ref{fig7}(a) shows the contour integral path in $k_\rho$-plane with increasing $\kappa$. From residue theorem, it is well-known that the integral value is given by the residue of the poles surrounded by the integration path. When $\kappa$ is small, corresponding to two real and distinct modes (red crosses in Fig.~\ref{fig7}(a)), the value of the integral changes in opposite directions as the upper limit passes the poles. The residue of the integrand has different signs for these two poles. However, when $\kappa$ hits the exceptional point (purple cross in Fig.~\ref{fig7}(a)), the value of the integral monotonically increases even after passing the pole. Entering the broken phase by increasing $\kappa$ (yellow crosses in Fig.~\ref{fig7}(a)), the increasing behavior could be preserved, but the integral values are substantially smaller. Since in this regime the poles move away from the real axis, the contribution from these poles decrease the integral values in the limit of $\kappa_\rho\rightarrow\infty$. This quantitative assessment agrees with the qualitative behavior of the poles deduced from the scattering parameters. More importantly, it reveals the physics underlying exceptional points on dipolar emission.

\section{Variation of reflection coefficients for TE-modes}
\begin{figure}
\includegraphics[scale=0.9]{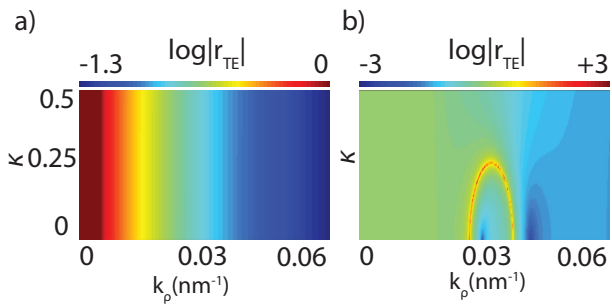}
\caption{\label{fig8} Reflection coefficient of TE-mode as a function of in-plane momentum and non-Hermiticy parameter at (a) E= 1.2 eV and (b) E= 3.6 eV.}
\end{figure}

In the main text, we described the reflection coefficient for TM modes at 1.2 eV. Figure~\ref{fig8}(a) shows the reflection coefficient of TE modes in $k_\rho\kappa$-plane. As described in the main text, the map is nearly featureless, since there are no TE modes at this energy. Accordingly, Purcell factors are very small. However, the structure supports TE modes at higher energies. Figure~\ref{fig8}(b) shows the reflection coefficient of TE modes at E = 3.6 eV in the $k_\rho\kappa$-plane. Note that a looping behavior occurs at $\kappa=0.26$, corresponding to the resonance feature of the vertical magnetic dipole as depicted in Fig.~\ref{fig3}(b). 
\bibliographystyle{apsrev4-1}
\bibliography{ref-PF}
\end{document}